\documentclass{article}
\usepackage[utf8]{inputenc}
\usepackage[margin=2.5cm]{geometry}
\usepackage{soul}
\usepackage[table,xcdraw]{xcolor}
\usepackage{graphicx}
\usepackage{multirow}
\usepackage{xspace}
\usepackage[most]{tcolorbox}
\usepackage{booktabs}
\usepackage{pifont}
\usepackage{balance}
\usepackage{amsmath,amsfonts}
\usepackage{amssymb}
\usepackage{svg}
\usepackage{url}
\usepackage{cryptocode}
\usepackage{booktabs}
\usepackage{graphicx}
\usepackage{subfig}
\usepackage{float}
\usepackage{circledsteps}
\usepackage{algpseudocode}
\usepackage{algorithm}
\usepackage{algcompatible}
\usepackage{multicol}

\usepackage[utf8]{inputenc}
\usepackage{hyperref}
\usepackage{graphicx}
\usepackage{xcolor}

\usepackage{amsmath}
\usepackage{amsthm}
\usepackage{amssymb}
\usepackage{amsfonts}
\usepackage{bbm}
\usepackage{algorithm}
\usepackage{algpseudocode}
\usepackage{mathtools}

\floatstyle{ruled}
\newfloat{algo}{htbp}{algo}
\floatname{algo}{Algorithm}

\usepackage{numbertabbing}

\usepackage{IEEEtrantools}
\allowdisplaybreaks

\usepackage[shortlabels]{enumitem}

\usepackage{tikz}
\usetikzlibrary{calc}
\usetikzlibrary{arrows}
\usetikzlibrary{arrows.meta}
\usetikzlibrary{patterns}
\usetikzlibrary{positioning}
\usetikzlibrary{decorations.pathreplacing}
\usetikzlibrary{shapes.misc}
\usetikzlibrary{spy}

\usepackage{pgfplots}
\pgfplotsset{compat=1.14}

\definecolor{myParula01Blue}{RGB}{0,114,189}
\definecolor{myParula02Orange}{RGB}{217,83,25}
\definecolor{myParula03Yellow}{RGB}{237,177,32}
\definecolor{myParula04Purple}{RGB}{126,47,142}
\definecolor{myParula05Green}{RGB}{119,172,48}
\definecolor{myParula06LightBlue}{RGB}{77,190,238}
\definecolor{myParula07Red}{RGB}{162,20,47}

\tikzset{myparula11/.style={color=myParula01Blue,solid,mark=+,mark options={solid}}}
\tikzset{myparula12/.style={color=myParula01Blue,densely dashed,mark=x,mark options={solid}}}
\tikzset{myparula13/.style={color=myParula01Blue,densely dotted,mark=o,mark options={solid}}}
\tikzset{myparula14/.style={color=myParula01Blue,dashdotted,mark=triangle,mark options={solid}}}
\tikzset{myparula15/.style={color=myParula01Blue,dashdotdotted,mark=square,mark options={solid}}}

\tikzset{myparula21/.style={color=myParula02Orange,solid,mark=+,mark options={solid}}}
\tikzset{myparula22/.style={color=myParula02Orange,densely dashed,mark=x,mark options={solid}}}
\tikzset{myparula23/.style={color=myParula02Orange,densely dotted,mark=o,mark options={solid}}}
\tikzset{myparula24/.style={color=myParula02Orange,dashdotted,mark=triangle,mark options={solid}}}
\tikzset{myparula25/.style={color=myParula02Orange,dashdotdotted,mark=square,mark options={solid}}}

\tikzset{myparula31/.style={color=myParula03Yellow,solid,mark=+,mark options={solid}}}
\tikzset{myparula32/.style={color=myParula03Yellow,densely dashed,mark=x,mark options={solid}}}
\tikzset{myparula33/.style={color=myParula03Yellow,densely dotted,mark=o,mark options={solid}}}
\tikzset{myparula34/.style={color=myParula03Yellow,dashdotted,mark=triangle,mark options={solid}}}
\tikzset{myparula35/.style={color=myParula03Yellow,dashdotdotted,mark=square,mark options={solid}}}

\tikzset{myparula41/.style={color=myParula04Purple,solid,mark=+,mark options={solid}}}
\tikzset{myparula42/.style={color=myParula04Purple,densely dashed,mark=x,mark options={solid}}}
\tikzset{myparula43/.style={color=myParula04Purple,densely dotted,mark=o,mark options={solid}}}
\tikzset{myparula44/.style={color=myParula04Purple,dashdotted,mark=triangle,mark options={solid}}}
\tikzset{myparula45/.style={color=myParula04Purple,dashdotdotted,mark=square,mark options={solid}}}

\tikzset{myparula51/.style={color=myParula05Green,solid,mark=+,mark options={solid}}}
\tikzset{myparula52/.style={color=myParula05Green,densely dashed,mark=x,mark options={solid}}}
\tikzset{myparula53/.style={color=myParula05Green,densely dotted,mark=o,mark options={solid}}}
\tikzset{myparula54/.style={color=myParula05Green,dashdotted,mark=triangle,mark options={solid}}}
\tikzset{myparula55/.style={color=myParula05Green,dashdotdotted,mark=square,mark options={solid}}}

\tikzset{myparula61/.style={color=myParula06LightBlue,solid,mark=+,mark options={solid}}}
\tikzset{myparula62/.style={color=myParula06LightBlue,densely dashed,mark=x,mark options={solid}}}
\tikzset{myparula63/.style={color=myParula06LightBlue,densely dotted,mark=o,mark options={solid}}}
\tikzset{myparula64/.style={color=myParula06LightBlue,dashdotted,mark=triangle,mark options={solid}}}
\tikzset{myparula65/.style={color=myParula06LightBlue,dashdotdotted,mark=square,mark options={solid}}}

\tikzset{myparula71/.style={color=myParula07Red,solid,mark=+,mark options={solid}}}
\tikzset{myparula72/.style={color=myParula07Red,densely dashed,mark=x,mark options={solid}}}
\tikzset{myparula73/.style={color=myParula07Red,densely dotted,mark=o,mark options={solid}}}
\tikzset{myparula74/.style={color=myParula07Red,dashdotted,mark=triangle,mark options={solid}}}
\tikzset{myparula75/.style={color=myParula07Red,dashdotdotted,mark=square,mark options={solid}}}

\theoremstyle{plain}
\newtheorem{theorem}{Theorem}
\newtheorem*{theorem*}{Theorem}

\newtheorem{lemma}{Lemma}

\theoremstyle{definition}
\newtheorem{definition}{Definition}

\usepackage{xspace}
\newcommand{\cf}[0]{cf.\xspace}
\newcommand{\ie}[0]{\emph{i.e.}\xspace}
\newcommand{\eg}[0]{\emph{e.g.}\xspace}

\tikzset{blockchain/.style={
        x=1.25cm,
        y=1.25cm,
        node distance=0.5cm,
        block/.style = {
            minimum width=0.75cm,
            minimum height=0.75cm,
            draw,
            shade,
            top color=white,
            bottom color=black!10,
        },
        block-adv/.style = {
            block,
            bottom color=myParula07Red!50,
            draw=myParula07Red!50!black,
        },
        block-hon/.style = {
            block,
            bottom color=myParula05Green!50,
            draw=myParula05Green!50!black,
        },
        link/.style = {
            -latex,
        },
        link-adv/.style = {
            link,
        },
        link-hon/.style = {
            link,
        },
    }
}

\usepackage{pifont}
\usepackage[flushleft]{threeparttable}

\newcommand{\Thorizon}[0]{\ensuremath{T_\mathsf{hor}}}
\newcommand{\chain}[0]{\ensuremath{\mathsf{ch}}}
\newcommand{\Chain}[0]{\ensuremath{\mathsf{Ch}}}
\newcommand{\chainava}[0]{\ensuremath{\mathsf{chAva}}}
\newcommand{\chainfin}[0]{\ensuremath{\mathsf{chFin}}}
\newcommand{\chainconf}[0]{\ensuremath{\mathsf{Ch}}}

\newcommand{\rlmdghost}[0]{\ensuremath{\operatorname{\textsc{RLMD-GHOST}}}}

\newcommand{\hfc}[0]{\ensuremath{\operatorname{\textsc{HFC}}}}

\newcommand{\ghost}[0]{\ensuremath{\operatorname{\textsc{GHOST}}}}

\newcommand{\C}[0]{\ensuremath{\mathcal{C}}}
\newcommand{\LJ}[0]{\ensuremath{\mathcal{LJ}}}

\DeclareMathOperator*{\argmax}{arg\,max}

\long\def\blockcomment#1\endblockcomment{}
\algdef{SE}[DOWHILE]{Do}{doWhile}{\algorithmicdo}[1]{\algorithmicwhile\ #1}%

\algdef{SE}[UPON]{Upon}{EndUpon}[1]{\textbf{upon}\ $#1$}{\textbf{end upon}}%
\algdef{SE}[AT]{At}{EndAt}[1]{\textbf{at}\ $#1$}{\textbf{end at}}%

\newcommand{\GAT}[0]{\ensuremath{\mathsf{GAT}}}
\newcommand{\GST}[0]{\ensuremath{\mathsf{GST}}}

\newcommand{\V}[0]{\ensuremath{\mathcal{V}}}

\newcommand{\FC}[0]{\ensuremath{\mathsf{FC}}}
\newcommand{\HFC}[0]{\ensuremath{\mathsf{HFC}}}
\newcommand{\negl}[0]{\ensuremath{\operatorname{negl}}}

\newcommand{\Tconf}[0]{\ensuremath{T_\mathsf{conf}}}
\newcommand{\Tafter}[0]{\ensuremath{T_\mathsf{sec}}}

\newcommand{\B}[0]{\ensuremath{\mathcal{B}}}

\newcommand{\node}[0]{\ensuremath{\mathcal{P}}}

\newcommand{\genesis}[0]{\ensuremath{B_\text{genesis}}\xspace}

\newcommand{\Goldfish}[0]{\textsf{Goldfish}\xspace}
\newcommand{\LMDGHOST}[0]{\textsf{LMD-GHOST}\xspace}
\newcommand{\RLMDGHOST}[0]{\textsf{RLMD-GHOST}\xspace}

\newcommand{\FIL}[0]{\textsf{FIL}\xspace}

\newcommand{\LOGbft}[2]{%
    \ifthenelse{\equal{#1}{}}{%
        \ensuremath{\mathsf{LOG}_{\mathrm{bft}}^{#2}}%
    }{%
        \ensuremath{\mathsf{LOG}_{\mathrm{bft},#1}^{#2}}%
    }%
}

\newcommand{\ld}[1]{%
    \ifthenelse{\equal{#1}{}}{%
        \ensuremath{\mathrm{L}^{(c)}}%
    }{%
        \ensuremath{\mathrm{L}^{(#1)}}%
    }%
}

\newcommand{\bprop}[1]{%
    \ifthenelse{\equal{#1}{}}{%
        \ensuremath{\Hat{b}}%
    }{%
        \ensuremath{\Hat{b}_{#1}}%
    }%
}

\title{A Simple Single Slot Finality Protocol For Ethereum}
\author{Francesco D'Amato\\
  Ethereum Foundation\\
  \url{francesco.damato@ethereum.org}
  \and Luca Zanolini\\
   Ethereum Foundation\\
  \url{luca.zanolini@ethereum.org}
}

\date{}

\begin{document}
\maketitle
\begin{abstract}\noindent

Currently, Gasper, the implemented consensus protocol of Ethereum, takes between 64 and 95 slots to finalize blocks. Because of that, a significant portion of the chain is susceptible to reorgs. The possibility to capture MEV (Maximum Extractable Value) through such reorgs can then disincentivize honestly following the protocol, breaking the desired correspondence of honest and rational behavior. Moreover, the relatively long time to finality forces users to choose between economic security and faster transaction confirmation. This motivates the study of the so-called single slot finality protocols: consensus protocols that finalize a block in each slot and, more importantly, that finalize the block proposed at a given slot within such slot. 

In this work we propose a \emph{simple, non-blackbox} protocol that combines a synchronous dynamically available protocol with a {partially synchronous} finality gadget, resulting in a consensus protocol that can finalize one block per slot, paving the way to \emph{single slot finality} within Ethereum. Importantly, the protocol we present can finalize the block proposed in a slot, within such slot. 

\end{abstract}
\section{Introduction}
\label{sec:introduction}
Traditional Byzantine consensus protocols, such as PBFT~\cite{DBLP:conf/osdi/CastroL99}, are devised in a partial synchronous network model~\cite{DBLP:journals/jacm/DworkLS88}, in the sense that they always guarantee safety, but they guarantee liveness only after $\GST$. In this setting, however, participants in the protocol are fixed, known in advance, and without possibility to go \emph{offline}. 

Dynamic participation (among systems' participants) has lately become an essential prerequisite for developing permissionless consensus protocols. This concept, initially formalized by Pass and Shi via their \emph{sleepy model},~\cite{sleepy} encapsulates the ability of a system to handle participants joining or leaving during a protocol execution. In particular, a consensus protocol that preserves safety and liveness while allowing dynamic participation is called \emph{dynamically available}. \\
\indent One problem of such protocols, as a result of the CAP theorem~\cite{DBLP:journals/sigact/GilbertL02}\cite{cap2}, is that they do not tolerate network partitions; no consensus protocols can both satisfy liveness (under dynamic participation) and safety (under temporary network partitions). Simply put, a consensus protocol (for state-machine replication) cannot produce a single chain that concurrently offers dynamic availability and guarantees transaction finality in case of asynchronous periods or network partitions. Because of that, dynamically available protocols studied so far are focused on a synchronous model~\cite{goldfish}\cite{DBLP:journals/iacr/MalkhiMR22}\cite{DBLP:conf/ccs/Momose022}. \\
\indent To overcome this impossibility result, Neu \emph{at al.}~\cite{DBLP:conf/sp/NeuTT21} introduce a family of protocols, referred to as \emph{ebb-and-flow} protocols, which operate under two confirmation rules, and outputting two chains, one a prefix of the other. The first confirmation rule defines what is known as the \emph{available chain}, which provides liveness under dynamic participation (and synchrony). The second confirmation rule defines the \emph{finalized chain}, and provides safety even under network partitions. Interestingly, such family of protocols also captures the nature of the Ethereum consensus protocol, Gasper~\cite{gasper}, in which the available chain is output by (the confirmation rule of) \LMDGHOST~\cite{zamfir}, and the finalized chain by the (confirmation rule of the) \emph{finality gadget} Casper~FFG~\cite{casper}. However, the (original version of) \LMDGHOST is actually not secure~\cite{DBLP:conf/sp/NeuTT21} even in a context of full-participation. \\
\indent Motivated by finding a (more secure) alternative to \LMDGHOST, and following the ebb-and-flow approach, D'Amato \emph{et al.}~\cite{goldfish} devise a synchronous dynamically available consensus protocol, \Goldfish, that, combined with a generic (partially synchronous) finality gadget, implements an ebb-and-flow protocol. Moreover, \Goldfish is reorg resilient: blocks proposed by honest validators are guaranteed inclusion in the chain. However, \Goldfish is brittle to temporary asynchrony~\cite{rlmd}, in the sense that even a single violation of the bound of network delay can lead to a catastrophic failure, jeopardizing the safety of \emph{any} previously confirmed block, resulting in a protocol that is not practically viable to replace \LMDGHOST in Ethereum. In other words, \Goldfish is not \emph{asynchrony resilient}.\\
\indent To cope with the problem of \Goldfish, D'Amato and Zanolini~\cite{rlmd} propose \RLMDGHOST, a provably secure synchronous protocol that does not lose safety during \emph{bounded} periods of asynchrony and which tolerates a weaker form of dynamic participation, offering a trade-off between dynamic availability and asynchrony resilience. Their protocol results appealing for practical systems, where strict synchrony assumptions might not always hold, contrary to what is generally assumed with standard synchronous protocols. \\
\indent In this work we build upon the work of D'Amato and Zanolini~\cite{rlmd}, and we devise a protocol that combines \RLMDGHOST with a {partially synchronous} finality gadget. In particular, we give the following contributions. We devise a secure and reorg-resilient ebb-and-flow protocol~\cite{DBLP:conf/sp/NeuTT21} as a potential substitute for the current Ethereum consensus protocol, Gasper~\cite{gasper}, which can finalize (at most) one block per slot. In particular, our protocol can finalize the block proposed in the current slot, within such slot, paving the way to \emph{single slot finality}~\cite{ssf} protocols for practical use within Ethereum. Finally, we expand upon the \emph{generalized sleepy model}~\cite{rlmd} introduced by D'Amato and Zanolini\cite{rlmd}, adjusting it to accommodate a partially synchronous setting. We refer to the resulting model as the \emph{generalized partially synchronous sleepy model}. This enhanced model not only extends the original sleepy model, first presented by Pass and Shi~\cite{sleepy}, but it also introduces stronger and more generalized constraints related to the corruption and sleepiness power of the adversary. Furthermore, our model integrates the concept of partial synchrony, setting it apart from the model proposed by D'Amato and Zanolini~\cite{rlmd}. Our security results will be proven within this extended model.
\indent The remainder of this work is structured as it follows. In Section~\ref{sec:model} we present our system model. Prerequisites for this work are presented in Section~\ref{sec:prerequisites}; we recall \RLMDGHOST as originally presented by D'Amato and Zanolini~\cite{rlmd}, state its properties, and show a class of protocols, called \emph{propose-vote-merge} protocols, that groups together (a variant of) \LMDGHOST, (a variant of) \Goldfish, and \RLMDGHOST under an unique framework. Protocol specifications are described in Section~\ref{sec:specific}. In particular, we show how to slightly modify \RLMDGHOST to interact with a finality gadget, and then present the full protocol. In Section~\ref{sec:analysis} we formally prove the properties that our protocol satisfy. Finally, in Section~\ref{sec:ssf} we enable our protocol to finalize the block proposed in the current slot through \emph{acknowledgments}, messages sent by participants in the consensus protocol, but only relevant to external observers. Conclusions are drawn in Section~\ref{sec:conclusions}. 

\section{Related works}
\label{sec:related-works}

Pass and Shi~\cite{sleepy} introduced the \emph{sleepy model of consensus}, which models a distributed system where the participants can be either online or offline, meaning their participation is dynamic. This differs from the standard models in the literature that assume honest participants are always online and execute the assigned protocol. Dynamic participation became a key requirement to devise consensus protocols, as it adds a more robustness to systems that allow participants to go offline, while preserving safety and
liveness of such \emph{dynamically available} protocols.

Neu et al~\cite{DBLP:conf/sp/NeuTT21} introduce the \emph{partially synchronous sleepy model} and define the objectives of the Ethereum consensus protocol, Gasper~\cite{gasper}, through the concept of an \emph{ebb-and-flow protocol}. A secure ebb-and-flow protocol produces both a dynamically available ledger and a finalized ledger, that is always safe and live after $\max\{\GST,\GAT\}$. In the context of Gasper, the dynamically available ledger is defined by \LMDGHOST~\cite{zamfir} and the finalized ledger by Casper~\cite{casper}.

However, under a deeper analysis, Neu \emph{et al}~\cite{DBLP:conf/sp/NeuTT21} show that \LMDGHOST is not dynamically available, by presenting an attack to its liveness.  D'Amato \emph{et al.}~\cite{goldfish} introduce \Goldfish, a simplified variant of \LMDGHOST, aiming at solving some problems related to \LMDGHOST~\cite{DBLP:conf/sp/NeuTT21, ethresearch-balancing-attack}, that results in a synchronous dynamically available protocol in the partially synchronous sleepy model that, composed with a generic finality gadget, implements an ebb-and-flow protocol. \Goldfish however is brittle to temporary asynchrony, in the sense that even a single violation of the bound of network delay can lead to a catastrophic failure, jeopardizing the safety of \emph{any} previously confirmed block.

D'Amato and Zanolini~\cite{rlmd} introduce the \emph{generalized sleepy model}. This model takes up from the original sleepy model presented by Pass and Shi~\cite{sleepy} and extends it with more generalized and stronger constraints in the corruption and sleepiness power of the adversary. This allow to explore a broad space of dynamic participation regimes which fall between complete dynamic participation and no dynamic participation. Moreover, they introduce \RLMDGHOST, a generalization of (variants of) \Goldfish and \LMDGHOST, that offers a trade-off between resilience to temporary asynchrony and dynamic availability. \RLMDGHOST represents a middle ground between \LMDGHOST, an asynchrony resilient but not dynamically available protocol, and \Goldfish, a dynamically available but not asynchrony resilient protocol. \RLMDGHOST is resilient to bounded asynchrony \emph{up to a vote expiry period}, and satisfies an appropriate notion of dynamic availability in the generalized sleepy model.

\section{Model and Preliminary Notions}
\label{sec:model}

\subsection{System model}

We consider a set of $n$ \emph{validators} $v_1, \dots, v_n$ that communicate with each other through exchanging messages. Every validator is identified by a unique cryptographic identity and the public keys are common knowledge. Validators are assigned a protocol to follow, consisting of a collection of programs with instructions for all validators. A validator that follows its protocol during an execution is called \emph{honest}. Each validator has a \emph{stake}, which we assume to be the same for every validator. {If a validator $v_i$ fails to serve the role assigned to it or tries to deliberately deviate from the protocol, i.e., $v_i$ is \emph{Byzantine}, and a proof of this misbehavior is given, it loses a part of its stake proportional to the severity of the fault ($v_i$ gets \emph{slashed})}. We assume the existence of a probabilistic poly-time adversary $\mathcal{A}$ that can choose up to $f$ validators to corrupt over an entire protocol execution. Corrupted validators stay corrupted for the remaining duration of the protocol execution, and are thereafter called \emph{adversarial}. The adversary $\mathcal{A}$ knows the the internal state of adversarial validators. The adversary is \emph{adaptive}: it chooses the corruption schedule dynamically, during the protocol execution. 

We assume that a best-effort gossip primitive that will reach all validators is available. In a protocol, this primitive is accessed through the events “sending a message through gossip” and “receiving a gossiped message.”  Moreover, we assume that messages from honest validator to honest validator are eventually received and cannot be forged. This includes messages sent by Byzantine validators, once they have been received by some honest validator $v_i$ and gossiped around by $v_i$. 

Time is divided into discrete \emph{rounds}. We consider a partially synchronous model in which validators have synchronized clocks but there is no a priori bound on message delays. However, there is a time (not known by the validators), called \emph{global stabilization time} (\GST), after which message delays are bounded by $\Delta$ rounds. Moreover, we define the notion of \emph{slot} as a collection of $4\Delta$ rounds. The adversary $\mathcal{A}$ can decide for each round which honest validator is \emph{awake} or \emph{asleep} at that round~\cite{sleepy}. Asleep validators do not execute the protocol and messages for that round are queued and delivered in the first round in which the validator is awake again. Honest validators that become awake at round $r$, before starting to participate in the protocol, must first execute (and terminate) a \emph{joining protocol} (Section~\ref{sec:prerequisites}), after which they become \emph{active}.
All adversarial validators are always awake, and are not prescribed to follow any protocol. Therefore, we always use active, awake, and asleep to refer to honest validators. As for corruptions, the adversary is adaptive also for sleepiness, \ie, the sleepiness schedule is also chosen dynamically by the adversary. Moreover, there is a time (not known by the validators), called \emph{global awake time} (\GAT), after which all validators are always~awake.

We assume that every message has an \emph{expiration period} $\eta$~\cite{goldfish}\cite{rlmd}. More specifically, for a given slot~$t$ and a constant $\eta \in \mathbb{N}$ greater than or equal to~$0$, the \emph{expiration period} for slot~$t$ is the interval $[t-\eta, t-1]$. Only messages sent within this time frame influence the behavior of the protocol at slot~$t$. Furthermore, during each protocol execution slot, only the most recent messages sent by validators are considered.

We require that, for some fixed parameter $1\leq \tau \leq \infty$, the following condition, referred by D'Amato and Zanolini~\cite{rlmd} as \emph{$\tau$-sleepiness at slot $t$}, holds for any slot $t$ \emph{after $\GST$}:
\begin{equation}
\label{eq:sleepy-req}
  |H_{t-1}| > |A_{t} \cup (H_{t-\tau, t-2}\setminus H_{t-1})|
\end{equation}

with $H_t$, $A_t$, and $H_{s, t}$ are the set of active validators at round $4\Delta t + \Delta$, the set of adversarial validators at round $4\Delta t + \Delta$, and the set of validators that are active \emph{at some point} in slots $[s,t]$, \ie, $H_{s,t} = \bigcup_{i=s}^t H_i$ (if $i < 0$ then $H_i \coloneqq \emptyset$), respectively. Note that $f = \lim_{t \to \infty} |A_t|$.  In other words, we require the number of active validators at round $4\Delta (t-1) + \Delta$ to be greater than the number of adversarial validators at round $4\Delta t + \Delta$, together with the number of validators that were active {at some point} between rounds $4\Delta (t-\tau) + \Delta$ and $4\Delta (t-2) + \Delta$, but not at round $4\Delta (t-1) + \Delta$. 

Intuitively, this condition is designed to work with a protocol that applies expiration to its messages, with the period set as $\eta = \tau$. The messages taken into consideration at slot~$t$ originate from slots $[t-\tau, t-1]$. Among these, the only messages sent by honest validators that can be relied upon come from $H_{t-1}$. However, unexpired messages from honest validators, who were inactive in slot $t-1$, could potentially aid the adversary.

Note that our approach diverges from the \emph{generalized sleepy model} proposed by D'Amato and Zanolini~\cite{rlmd}. Specifically, we require that Equation~\ref{eq:sleepy-req} only holds after $\GST$ and we refer to this model as the \emph{generalized partially synchronous $\tau$-sleepy model} (or wlog, when the context is clear, as the \emph{$\tau$-sleepy model} for short). Finally, we say that an execution in the generalized partially synchronous sleepy model is \emph{$\tau$-compliant} if it satisfies $\tau$-sleepiness (Equation~\ref{eq:sleepy-req}).

\subsection{Validator internals}

\paragraph*{View.} A \emph{view} (at a given round $r$), denoted by $\V$, is a subset of all the messages that a validator has received until $r$. The notion of view is \emph{local} for the validators. For this reason, when we want to focus the attention on a specific view of a validator $v_i$, we denote with $\V_i$ the view of $v_i$ (at a round $r$).
\paragraph*{Blocks and chains.} Let's consider two chains, $\chain_1$ and $\chain_2$. We denote $\chain_1 \prec \chain_2$ if $\chain_1$ acts as a prefix to $\chain_2$. When block $B$ is at the end of chain $\chain$, we refer to it as the \emph{head of $\chain$}, and we equate the entire chain with~$B$. Therefore, if $\chain' \prec \chain$ and $A$ is the head of $\chain'$, we also express this as $\chain' \prec B$ and $A \prec B$.
\paragraph{Fork-choice functions.} A \emph{fork-choice function} is a deterministic function, denoted as $\FC$. This function, when given a view $\V$ and a slot $t$ as inputs, produces a block $B$. If~$B$ is a block extending~$\FC(\V, t)$, then $\FC(\V \cup {B}, t)$ equals~$B$. The result of $\FC$ is referred to as the \emph{head of the canonical chain} in $\V$, and the chain with $B$ as its head is referred to as the \emph{canonical chain} in $\V$. Every validator keeps track of its canonical chain and updates it using $\FC$, according to its local view. The canonical chain for validator $v_i$ at round~$r$ is represented as~$\chain_i^r$. In this work we will focus our attention on a specific class of fork-choice functions based on $\ghost$~\cite{ghost}. D'Amato and Zanolini~\cite{rlmd} characterize a $\ghost$-based fork-choice function by a view filter $\FIL$, which takes as input a view $\V$ and a slot $t$, and outputs $(\V', t)$, where $\V'$ is another view such that $\V' \subseteq \V$. Then, $\FC(\V, t) \coloneqq \ghost(\FIL(\V, t))$, i.e., $\FC \coloneqq \ghost \circ\,\FIL$.

\subsection{Security}

\paragraph*{Security Parameters.}
In this work we treat $\lambda$ and $\kappa$ as the security parameters related to the cryptographic components utilized by the protocol and the protocol's own security parameter, respectively. We also account for a finite time horizon, represented as $\Thorizon$, which is polynomial in relation to $\kappa$. An event is said to occur with \emph{overwhelming probability} if it happens except with probability which is $\negl(\kappa) + \negl(\lambda)$. The properties of cryptographic primitives hold true with a probability of $\negl(\lambda)$, signifying an overwhelming probability, although we will not explicitly mention this in the subsequent sections of this work.

\paragraph*{Confirmed chain.}
The protocols we consider always specify a \emph{confirmation rule}, with whom validators can identify a \emph{confirmed prefix} of the canonical chain. Alongside the canonical chain, validators then also keep track of a \emph{confirmed chain}. We refer to the confirmed chain of validator~$v_i$ at round~$r$ as~$\chainconf_i^r$ (\cf $\chain_i^r$ for the canonical chain). 

\begin{definition}[Secure protocol~\cite{goldfish}]
 \label{def:security}
We say that a protocol outputting a confirmed chain $\Chain$ is \emph{secure} after time $\Tafter$, and has confirmation time $\Tconf$\footnote{If the protocol satisfies liveness, then at least one honest proposal is added to the confirmed chain of all active validators every $\Tconf$ slots. Since honest validators include all transactions they see, this ensures that transactions are confirmed within time $\Tconf + \Delta$ (assuming infinite block sizes or manageable transaction volume).}, if $\Chain$ satisfies:
    \begin{description}
        \item[Safety] For any two rounds $r, r' \geq \Tafter$, and any two honest validators $v_i$ and $v_j$ (possibly $i=j$) at rounds $r$ and $r'$ respectively, either $\Chain_i^r \prec \Chain_{j}^{r'}$ or $\Chain_j^{r'} \prec \Chain_i^r$.

        \item[Liveness] For any rounds $r \geq \Tafter$ and $r' \geq r+\Tconf$, and any honest validator~$v_i$ active at round $r'$, $\Chain_{i}^{r'}$ contains a block proposed by an honest validator at a round $> r$.
    \end{description}
A protocol satisfies \emph{$\tau$-safety} and \emph{$\tau$-liveness} if it satisfies safety and liveness, respectively, \emph{in the $\tau$-sleepy model}, \ie, in $\tau$-compliant executions. A protocol satisfies $\tau$-security if it satisfies $\tau$-safety and $\tau$-liveness. 
\end{definition}

We now recall the definitions of \emph{dynamic availability} and \emph{reorg resilience} from~\cite{rlmd}. We consider them only under network synchrony, \ie, for $\GST = 0$, as this is the only setting in which we utilize them. Note that it is customary to only analyze dynamic availability with $\GST = 0$, when analyzing the behavior of ebb-and-flow protocols.

\begin{definition}[Dynamic availability]
 \label{def:dyn-ava}
We say that a protocol is $\tau$-\emph{dynamically-available} if and only if it satisfies $\tau$-security with confirmation time $\Tconf = O(\kappa)$ when $\GST = 0$. Moreover, we say that a protocol is dynamically available if it is $1$-dynamically-available, as this corresponds to the usual notion of dynamic availability.
\end{definition}

\begin{definition}[Reorg resilience]
\label{def:reorg-resilience}
An execution with $\GST = 0$ satisfies \emph{reorg resilience} if any honest proposal $B$ from a slot $t$ is always in the canonical chain of all active validators at rounds $\geq 4\Delta t + \Delta$.
A protocol is \emph{$\tau$-reorg-resilient} if all $\tau$-compliant executions with $\GST = 0$ satisfy reorg resilience.
\end{definition}

\begin{definition}[Accountable safety]
    \label{def:acc-safety}
    We say that a protocol has \emph{accountable safety} with resilience $f > 0$ if, upon a safety violation, it is possible to identify at least $f$ responsible participants. In particular, it is possible to collect evidence from sufficiently many honest participants and generate a cryptographic proof that identifies $f$ adversarial participants as protocol violators. Such proof cannot falsely accuse any honest participant that followed the protocol correctly. Finally, we also say that a chain is $f$-\emph{accountable} if the protocol outputting it has accountable safety with resilience $f$. If a protocol $\Pi$ outputs multiple chains $\Chain_1, \dots, \Chain_k$, we say that $\Chain_i$ is $f$-accountable if $\Pi_i$ is, where $\Pi_i$ is the protocol which runs $\Pi$ and outputs only $\Chain_i$.
\end{definition}

\paragraph*{Ebb-and-flow protocols.}

Neu \emph{et al.}~\cite{DBLP:conf/sp/NeuTT21} propose a protocol with two confirmation rules that outputs two chains, one that provides liveness under dynamic participation (and synchrony), and one that provides accountable safety even under network partitions. This protocol is called \emph{ebb-and-flow} protocol. We present a generalization of it, in the $\tau$-sleepy model.

\begin{definition}[$\tau$-secure ebb-and-flow protocol]

A $\tau$-secure \emph{ebb-and-flow protocol} outputs an available chain $\chainava$ that is $\tau$-dynamically-available if $\GST=0$, and a finalized (and accountable) chain $\chainfin$ that, if $f<\frac{n}{3},$ is always safe and is live after $\max\{\GST,\GAT\}$. Moreover, for each honest validator $v_i$ and for every round $r$, $\chainfin_i^r$ is a prefix of $\chainava_i^r$.

\end{definition}

\section{Propose-vote-merge protocols}
\label{sec:prerequisites}

The aim of this work is to present a secure ebb-and-flow~\cite{DBLP:conf/sp/NeuTT21} protocol that can finalize (at most) one block per slot and, in particular, that can finalize within slot $t$ the block proposed in $t$. This is achieved by revisiting the \emph{propose-vote-merge} protocol \RLMDGHOST introduced by D'Amato and Zanolini~\cite{rlmd} as the basis for our protocol implementation. Propose-vote-merge protocols proceed in \emph{slots} consisting of $k$ rounds\footnote{D'Amato and Zanolini~\cite{rlmd} implement \RLMDGHOST with fast confirmation with $k=3\Delta$ (Appendix B~\cite{rlmd}). However, we will consider $k=4\Delta$, following the approach taken by D'Amato \emph{et al.}~\cite{goldfish} when presenting \Goldfish with \emph{fast confirmation}. We will show how \RLMDGHOST with fast confirmation can be changed into its variant with $k=4\Delta$ in Section~\ref{sec:specific} while presenting our protocol.}, each having a proposer~$v_p$, chosen through a proposer selection mechanism among the set of validators. In particular, at the beginning of each slot $t$, the proposer $v_p$ proposes a block $B$. Then, all active validators (also referred as \emph{voters}) vote after $\Delta$ rounds. Every validator $v_i$ has a buffer $\B_i$, a collection of messages received from other validators, and a view $\V_i$, used to make consensus decisions, which admits messages from the buffer only at specific points in time. 

Propose-vote-merge protocols are defined through a deterministic fork-choice function $\FC$, which is used by honest proposers and voters to decide how to propose and vote, respectively, based on their view at the round in which they are performing those actions. It is moreover used as the basis of a \emph{confirmation rule} (Section~\ref{sec:confirmation}), which defines the output of the protocol, and thus with respect to which the security of the protocol is defined. In the case of \RLMDGHOST, its fork-choice function $\rlmdghost$ considers the last (non equivocating) messages sent by validators that are not older than $t-\eta$ slots (for an expiration period~$\eta$), in order to make protocol's decisions. In particular, the filter function $\FIL_{\text{rlmd}}(\V, t)$ removes \emph{all but the latest messages within the expiry period $[t-\eta, t)$ for slot~$t$, from non-equivocating validators}, i.e., $\FIL_{\text{rlmd}} = \FIL_{\text{lmd}}\circ\FIL_{\eta\text{-exp}}\circ\FIL_{eq}$. Here, $\FIL_{\text{lmd}}(\V, t)$ removes all but the latest votes of every validator (possibly more than one) from $\V$ and outputs the resulting view, i.e., it implements the \emph{latest message} (LMD) rule, $\FIL_{\eta\text{-exp}}(\V, t)$ removes all votes from slots $< t-\eta$ from $\V$ and outputs the resulting view, and $\FIL_{eq}(\V, t)$ removes all votes by \emph{equivocating validators in $\V$}~\cite{equiv-disc}, i.e., validators for which $\V$ contains multiple, equivocating, votes for some slot $t$. 

A propose-vote-merge protocol proceeds in three phases:

\textsc{propose}: In this phase, which starts at the beginning of a slot, the proposer~$v_p$ merges its view $\V_p$ with its buffer $\B_p$, \ie, $\V_p \gets \V_p \cup \B_p$, and sets $\B_p \gets \emptyset$. Then, $v_p$ runs the fork-choice function $\FC$ with inputs its view $\V_p$ and slot $t$, obtaining the head of the chain $B' = \FC(\V_p, t)$. Proposer $v_p$ extends $B'$ with a new block $B$, and updates its canonical chain accordingly, setting $\chain_p \gets B$. Finally, it broadcasts the message [\textsc{propose}, $B$, $\V_p\,\cup \{B\}$, $t$, $v_p$].

\textsc{vote}: Here, every validator $v_i$ that receives a proposal message [\textsc{propose}, $B$, $\V$, $t$, $v_p$] from $v_p$ merges its view with the proposed view $\V$, by setting $\V_i \gets \V_i \cup \V$. Then, it broadcasts votes for some blocks based on its view. We omit, for the moment, for which blocks a validator $v_i$ votes: it will become clear once we present the full protocol. 

\textsc{merge:} In this phase, every validator $v_i$ merges its view with its buffer, \ie, $\V_i \gets \V_i \cup \B_i$, and sets $\B_i \gets \emptyset$.

The \textsc{merge} phase, along with all other operations involving views and buffers discussed in the previous section, are implemented using the \emph{view-merge} technique~\cite{goldfish}\cite{rlmd}\cite{highway}. The idea behind the view-merge technique involves synchronizing the views of all honest validators with the view $\V_p$ of the proposer for a specific slot \emph{before} the validators broadcast their votes in that slot.

D'Amato \emph{et al.}~\cite{goldfish} introduce the notion of \emph{active} validators\footnote{Observe that D'Amato \emph{et al.}~\cite{goldfish} actually refer to \emph{awake} validators to indicate what we call active, and to \emph{dreamy} validators to indicate what we call awake (but not active).}: awake validators that have terminated a \emph{joining protocol} at a round $r$, described as it follows. Assuming a propose-vote-merge protocol proceeding in {slots} of $k=4\Delta$ rounds, when an honest validator $v_i$ wakes up at some round $r \in (4\Delta (t-1) + 3\Delta, 4\Delta t + 3\Delta]$, it immediately receives all the messages that were sent while it was asleep, and it adds them into its buffer $\B_i$, without actively participating in the protocol yet. All new messages which are received are also added to the buffer $\B_i$. Validator $v_i$ then waits for the \emph{next view-merge opportunity}, at round $4\Delta t + 3\Delta$, in order to merge its buffer $B_i$ into its view $\V_i$. At this point, $v_i$ starts executing the protocol. From this point on, validator $v_i$ becomes \emph{active}, until either corrupted or put to sleep by the adversary. We consider such a joining protocol when describing our propose-vote-merge protocol.

\section{Protocol specification}
\label{sec:specific}

\subsection{Data structures}

We consider five message types: \textsc{propose}, \textsc{block}, \textsc{checkpoint}, \textsc{head-vote}, and \textsc{ffg-vote}. We make no distinctions between network-level representation of blocks and votes, and their representation in a validator's view, \ie, there is no difference between \textsc{block} and \textsc{*-vote} messages and blocks and votes, and we usually just refer to the latter. We describe the objects as tuples (\textsc{data-type}, $\dots$) with their data type as a tag, but in practice mostly refer to them without the tag. We use dot notation to refer to the fields. For the tag, we do so simply with $.\text{tag}$, for the other fields we use the generic names specified in the object descriptions below, to access the different fields, \eg, $B.t$ is the slot of block $B$. In the following, $t$ is a slot and $v_i$ a validator. 

\paragraph*{Blocks and checkpoints.}
A block is a tuple $B = (\textsc{block}, b, t, v_i)$, where $b$ is a \emph{block body}, \ie, the protocol-specific content of the block\footnote{For simplicity, we omit a reference to the parent block.}. A checkpoint is a tuple $\C = (\textsc{checkpoint}, B, t)$, where $B$ is a block and $\C.t \geq B.t$.

\paragraph*{Votes.}
A head vote is a tuple [\textsc{head-vote}, $B$, $t$, $v_i$], where $B$ is a block. An FFG vote is a tuple [\textsc{ffg-vote}, $\C_1$, $\C_2$, $v_i$], where $\C_1, \C_2$ are checkpoints, $\C_1.t < \C_2.t$, and $\C_1.B \prec \C_2.B$. We refer to the two checkpoints as \emph{source} and \emph{target}, respectively, and to FFG votes as \emph{links} between source and target. When $v_i$ is clear from context, we also write $\C_1 \to \C_2$ for the whole vote, \eg, we say that $v_i$ \emph{casts} a $\C_1 \to \C_2$ vote.

\paragraph*{Proposals.}
A proposal is a tuple [\textsc{propose}, $B$, $\V$, $t$, $v_i$] where $B$ is a block and $\V$ a view. We refer to $\V$ as a \emph{proposed view}.

\paragraph*{Gossip behavior.}
Votes and blocks are gossiped at any time, regardless of whether they are received directly or as part of another message. For example, a validator receiving a vote also gossips the block that it contains, and a validator receiving a proposal also gossips the blocks and votes contained in the proposed view. Finally, a proposal from slot $t$ is gossiped only during the first $\Delta$ rounds of slot~$t$.

\subsection{Confirmation rule}
\label{sec:confirmation}
A confirmation rule allows validators to identify a \emph{confirmed prefix} of the canonical chain, for which safety properties hold, and which is therefore used to define the output of the protocol. Since the protocol we are going to present outputs two chains, the available chain $\chainava$ and the finalized chain $\chainfin$, we have two confirmation rules. One is \emph{finality}, which we introduce in Section~\ref{sec:ffg}, and defines $\chainfin$. The other confirmation rule, defining $\chainava$, is the one adopted by \RLMDGHOST, in its variant supporting fast confirmation\footnote{With some minor changes, as \RLMDGHOST still has $3\Delta$ rounds per slots, by requiring an optimistic assumption on network latency in order for fast confirmations to be live.}. It is itself essentially split in two rules, a \emph{slow} $\kappa$-deep confirmation rule, which is live also under dynamic participation, and a \emph{fast optimistic rule}, requiring $\frac{2}{3}n$ honest validators to be awake, \ie, a stronger assumption than just $\tau-$compliance. Both rules are employed at round $4\Delta t +   2\Delta$, and $\chainava$ is updated to the highest block confirmed by either one, so that liveness of $\chainava$ only necessitates liveness of one of the two rules. In particular, $\tau$-compliance is sufficient for liveness. On the other end, safety of $\chainava$ requires both rules to be safe.

\subsection{FFG component}
\label{sec:ffg}

As mentioned above, a propose-vote-merge protocol is characterized by a fork-choice function that identifies for every slot the current head of the canonical chain for a given validator. Moreover, we described two kind of votes that a validator $v_i$ executes in the \textsc{vote} phase: a \textsc{head-vote}, used to vote for the head of the canonical chain, i.e., the output of the fork-choice function evaluated at the current slot, and an \textsc{ffg-vote}, used by the \emph{FFG-component} of our protocol\footnote{The component of our protocol that outputs $\chainfin$ is almost identical to Casper~\cite{casper}, the \emph{friendly} finality gadget (FFG) adopted by the Ethereum consensus protocol Gasper~\cite{gasper}. This is the reason why we decided to use the \emph{FFG} terminology already accepted within the Ethereum ecosystem.}.

The FFG component of our protocol aims at finalizing one block per slot by counting \textsc{ffg-votes} cast at a given slot.

\paragraph*{Justification.}
We say that a set of $\frac{2}{3}n$ distinct FFG votes $\C_1 \to \C_2$ is a \emph{supermajority link} between $\C_1$ and $\C_2$. We say that a checkpoint $C$ is \emph{justified} if there is a chain of $k \geq 0$ supermajority links (\genesis, $0) \to \C_1 \dots \to \C_{k-1} \to C$.  In particular, $(\genesis, 0)$ is justified. Finally, we say that a block $B$ \emph{is justified} if there exists a justified checkpoint $\C$ with $\C.B = B$.

\paragraph*{Slashing.}
The slashing rules are the same as in Casper~FFG. Validator $v_i$ is slashable (see Section~\ref{sec:model}) for two \emph{distinct} FFG votes $(\C_1, \C_2, v_i)$ and $(\C_3$, $\C_4, v_i)$ if either: {$\mathbf{E_1}$ (Equivocation)} $\C_2.t = \C_4.t$ or {$\mathbf{E_2}$ (Surround voting)} $\C_3.t < \C_1.t < \C_2.t < \C_4.t$.

\paragraph*{Latest justified checkpoint and block.}
A checkpoint is justified in a view $\V$ if $\V$ contains the chain of supermajority links justifying it. We refer to the justified checkpoint $\C$ of highest slot $\C.t$ in $\V$ as the \emph{latest justified checkpoint} in $\V$, or $\LJ(\V)$, and to $\LJ(\V).B$ as the \emph{latest justified block} in $\V$, or $LJ(\V)$. Ties are broken arbitrarily (the occurrence of a tie implies that $\frac{n}{3}$ validators are slashable for equivocation). For brevity, we also use $\LJ_i$ to refer to $\LJ(\V_i)$, the latest justified checkpoint in the view $\V_i$ of validator $v_i$.

\paragraph*{Finality.}
A checkpoint $\C$ is \emph{finalized} if it is justified and there exists a supermajority link $\C \to \C'$ with $\C'.t = \C.t + 1$. A block $B$ is finalized if there exists a finalized checkpoint $\C$ with $B = \C.B$.

\subsection{Voting}

\paragraph*{Fork-choice.}
Similarly to Gasper~\cite{gasper}, we adopt an hybrid \emph{justification-respecting} fork-choice, namely $\hfc$, building upon $\rlmdghost$~\cite{rlmd} fork-choice function. $\hfc(\V, t)$ starts from $LJ(\V)$, the \emph{latest justified block} in $\V$, instead of \genesis, and then proceeds as $\rlmdghost$, \ie, it runs $\ghost$ using the view filtered by $\FIL_{\text{rlmd}}$. Formally, we can define it by using another view filter, $\FIL_{\text{FFG}}$, \ie, $\hfc = \rlmdghost \circ\,\FIL_{\text{FFG}}$. $\FIL_{\text{FFG}}(\V, t)$ outputs $(\V', t)$, where $\V'$ filters out blocks in $\V$ that conflict with $LJ(\V)$. In other words, it filters out \emph{branches which do not contain $LJ(\V)$}, so $LJ(\V)$ is guaranteed to be canonical. 
\begin{algo}[h!]
\vbox{
\small
\begin{numbertabbing}\reset
  xxxx\=xxxx\=xxxx\=xxxx\=xxxx\=xxxx\=MMMMMMMMMMMMMMMMMMM\=\kill
\textbf{function} $\hfc(\V, t)$ \label{}\\
      \> \textbf{return} $\rlmdghost(\FIL_{\text{FFG}}(\V, t))$ \label{}\\
\textbf{function} $\FIL_{\text{FFG}}(\V, t)$ \label{}\\
    \> $\V' \gets \V\setminus\{B \in \V, B.\text{tag} = \textsc{block}: LJ(\V) \not \prec B \land B \not \prec LJ(\V)\}$\label{}\\
      \> \textbf{return} $(\V', t)$ \label{}\\[-5ex]
\end{numbertabbing}
}
\caption{$\hfc$, the justification-respecting fork-choice function}
\label{alg:hfc}
\end{algo}
\paragraph*{Voting rules.}
\label{para:voting-rules}
Consider a validator $v_i$ voting at slot $t$.
Head votes work exactly as in \RLMDGHOST, or any propose-vote-merge protocol, \ie, validators vote for the output of their fork-choice: when it is time to vote, validator $v_i$ casts vote [\textsc{head-vote}, $\hfc(\V_i, t), t, v_i$].
FFG votes always use the \emph{latest justified checkpoint as source}. The target block is the \emph{highest confirmed descendant of the latest justified block, or the latest justified block itself if there is none}. 
The target checkpoint is then $\C_{\text{target}} = (\argmax_{B \in \{LJ_i, \chainava\}}|B|,t)$, with $|B|$ being the height of block $B$, and the FFG vote of $v_i$ is [\textsc{ffg-vote}, $\LJ_i, \C_{\text{target}}, v_i$], voting for the link $\LJ_i \to \C_{\text{target}}$. 

\subsection{Protocol execution}

Our protocol is implemented in Algorithm~\ref{alg:pvm-generic} and it works as it follows. Note that the \textsc{Propose} and \textsc{Head-vote} phases are \emph{exactly} as in a generic propose-vote-merge protocol (see Section~\ref{sec:prerequisites}). Moreover, a slot $t$ in our protocol begins at round $4\Delta t$. At any time, the finalized chain $\chainfin_i$ of validator $v_i$ just consists of the finalized blocks according to its view $\V_i$, so we omit explicit updates to $\chainfin$ in the following.

\begin{algo}[htb!]
\small
\begin{numbertabbing}\reset

xxxx\=xxxx\=xxxx\=xxxx\=xxxx\=xxxx\=MMMMMMMMMMMMMMMMMMM\=\kill
  \textbf{State} \label{}\\
  \> \(\V_i \gets \{\B_{\text{genesis}}\} \): view of validator $v_i$ \label{}\\
  \> \(\B_i \gets \emptyset \): buffer of validator $v_i$  \label{}\\
  \> \(\chain_i \gets B_{\text{genesis}}\): canonical chain of validator $v_i$ \label{}\\  
  \> \(t \gets 0\): the current slot \label{}\\
  \> \(r \gets 0\): the current round \label{}\\
  \textsc{propose}\\
  \textbf{at} $r=4\Delta t$ \textbf{do} \label{} \\
  \> \textbf{if} $v_i = v_p^t$ \textbf{then} \label{}\\
  \>\>$\V_i \gets \V_i \cup \B_i$, $\B_i \gets \emptyset$ , $B' \gets \hfc(\V_i, t)$ \label{line:alg2-fc1}\\
  \>\> $B \gets \mathsf{NewBlock}(B')$, $\chain_i \gets B$ \label{}\\
  \>\> send message [\textsc{propose}, $B$, $\V_i\,\cup \{B\}$, $t$, $v_i$] through gossip \label{}\\
     \textsc{Head-vote}\\
     \textbf{at} $r=4\Delta t + \Delta$ \textbf{do} \label{}\\
     \> $\chain_i \gets \hfc(\V_i, t)$ \label{line:alg2-fc2}\\
     \> send message [\textsc{head-vote}, $\hfc(\V_i, t)$, $t$, $v_i$] through gossip \label{line:alg2-fc3}\\
    \textsc{Confirm and ffg-vote}\\
     \textbf{at} $r=4\Delta t + 2\Delta$ \textbf{do} \label{}\\
     \> $B_{\text{fast}} \gets \genesis$ \label{} \\
     \> $S_{\text{fast}} \gets \{B \prec \chain_i\colon |\{v_j\colon \exists B' \succ B : \ [\textsc{head-vote}, B', t, v_j] \in \B_i \}| \geq \frac{2}{3}n\}$ \label{} \\
     \> \textbf{if} $S_{\text{fast}} \neq \emptyset$ \textbf{then}: \label{}\\
    \>\> $B_{\text{fast}} \gets \underset{S_{\text{fast}}}{\text{arg max}} |B|$ \label{}\\
    \> \textbf{if} $\neg (B_{\text{fast}} \prec  \chainava_i \land \chain_i^{\lceil \kappa} \prec \chainava_i) $ \textbf{then}: \label{}\\
     \>\> $\chainava_i \gets \underset{\chain \in \{\chain_i^{\lceil \kappa}, B_{\text{fast}}\}}{\text{arg max}} |\chain|$ \label{}\\
    \>  $\C_{\text{target}} \gets (\underset{B \in \{LJ_i, \chainava_i\}}{\text{arg max}}|B|,t)$\label{line:set-target-checkpoint}\\
     \> send message [\textsc{ffg-vote}, $\LJ_i$, $\C_{\text{target}}$, $v_i$] through gossip  \label{line:cast-ffg-vote}\\
    \textsc{merge}\\
     \textbf{at} $r=4\Delta t + 3\Delta$ \textbf{do} \label{}\\
     \> $\V_i \gets \V_i \cup \B_i$ \label{}\\
       \> $\B_i \gets \emptyset$ \label{}\\
    \textbf{upon} receiving a gossiped message
    [\textsc{propose}, $B$, $\V$, $t$, $v_p^t$] \textbf{do} \label{}\\
    \> $\B_i \gets \B_i \cup \{B\}$ \label{}\\  
    \> \textbf{if} $r \in [4\Delta t, 4\Delta t + \Delta]$ \textbf{then} \label{}\\
    \>\> $\V_i \gets \V_i \cup \V$ \label{}\\
    \textbf{upon} receiving a gossiped \textsc{block} $B$ \textbf{or} a gossiped \textsc{*-vote} $V$ from \(v_j\) \textbf{do} \label{}\\
    \> $\B_i \gets \B_i \cup \{B\}$ \textbf{or} $\B_i \gets \B_i \cup \{V\}$ \label{}\\
    [-5ex]
\end{numbertabbing}
\caption{Single slot finality protocol -- code for validator $v_i$}
\label{alg:pvm-generic}
\end{algo}

\textsc{Propose}: At round $4\Delta t$, proposer $v_p$ merges its view $\V_p$ with its buffer $\B_p$, \ie, $\V_p \gets \V_p \cup \B_p$, and sets $\B_p \gets \emptyset$. Then, $v_p$ runs the fork-choice function $\hfc$ with inputs its view $\V_p$ and slot $t$, obtaining the head of the chain $B' = \hfc(\V_p, t)$. Proposer $v_p$ extends $B'$ with a new block $B$, and updates its canonical chain accordingly, by setting $\chain_p \gets B$. Finally, it broadcasts the proposal [\textsc{propose}, $B$, $\V_p\,\cup \{B\}$, $t$, $v_p$].

\textsc{Head-vote}: In rounds $[4\Delta t, 4\Delta t + \Delta]$, a validator $v_i$, upon receiving a proposal message (\textsc{propose}, $B$, $\V$, $t$, $v_p$) from $v_p$, merges its view with the proposed view $\V$ by setting $\V_i \gets \V_i \cup \V$. At round $4\Delta t + \Delta$, even if no proposal is received, validator $v_i$ updates its canonical chain by setting $\chain_i \gets \hfc(\V_i, t)$, and casts the head vote (\textsc{head-vote}, $\hfc(\V_i, t)$, $t$, $v_i$). 

\textsc{Confirm}: At round $4\Delta t + 2\Delta$, a validator $v_i$ selects for fast confirmation the highest \emph{canonical} block $B_{\text{fast}} \prec \chain_i$  such that $\B_i$ contains $\geq \frac{2}{3}n$ votes from slot~$t$ for descendants of $B_{\text{fast}}$, from distinct validators. It then updates its confirmed chain $\chainava_i$ to the highest between $B_{\text{fast}}$ and $\chain_i^{\lceil \kappa}$, the $\kappa$-deep prefix of its canonical chain, \emph{as long as this does not result in updating $\chainava_i$ to some prefix of it} (we do not needlessly revert confirmations).

\textsc{ffg-vote}: At round $4\Delta t + 2\Delta$, after updating $\chainava_i$, a validator $v_i$ casts the FFG vote (\textsc{ffg-vote}, $\LJ_i$, $\C_{\text{target}}$, $v_i$), where $\C_{\text{target}} = (\underset{B \in \{LJ_i, \chainava_i\}}{\text{arg max}}|B|,t)$

\textsc{Merge}: At round $4\Delta t + 3\Delta$, every validator $v_i$ merges its view with its buffer, \ie, $\V_i \gets \V_i \cup \B_i$, and sets $\B_i \gets \emptyset$.

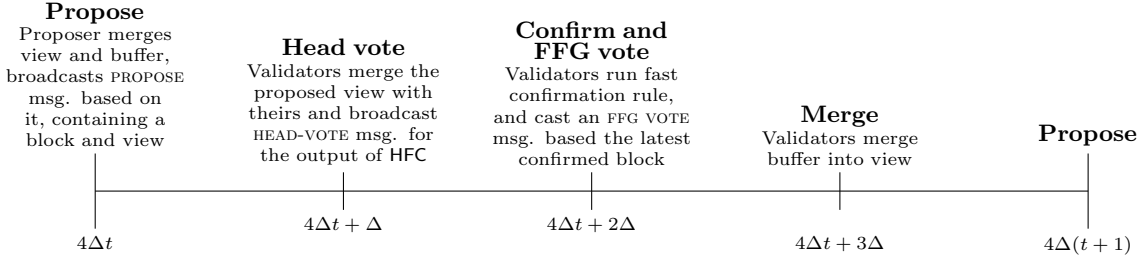
\begin{figure}
    \centering
\begin{tikzpicture}
\scriptsize
\pgfmathsetmacro{\segwidth}{3.3}

\draw (0,0) -- (4*\segwidth,0);

\draw (0,-0.5) -- (0,0.5);
\draw (\segwidth,-0.25) -- (\segwidth,0.25);
\draw (2*\segwidth,-0.25) -- (2*\segwidth,0.25);
\draw (3*\segwidth,-0.25) -- (3*\segwidth,0.25);
\draw (4*\segwidth,-0.5) -- (4*\segwidth,0.5);

\node[above, align=center, text width=2.7cm] at (0,0.5) {{\small \textbf{Propose}} \\ Proposer merges view and buffer, broadcasts \textsc{propose} msg. based on it, containing a block and view};
\node[above, align=center, text width=2.7cm] at (\segwidth,0.25) {{\small \textbf{Head vote}} \\ Validators merge the proposed view with theirs and broadcast \textsc{head-vote} msg. for the output of $\HFC$};
\node[above, align=center, text width=2.7cm] at (2*\segwidth,0.25) {{\small \textbf{Confirm and FFG vote}} \\ Validators run fast confirmation rule, and cast an \textsc{ffg vote} msg. based the latest confirmed block};
\node[above, align=center, text width=2.3cm] at (3*\segwidth,0.25) {{\small \textbf{Merge}} \\ Validators merge buffer into view};
\node[above, align=center, text width=2.3cm] at (4*\segwidth,0.5) {{\small \textbf{Propose}}};

\node[below, align=center, text width=2cm] at (0,-0.5) {\textbf{$4\Delta t$}};
\node[below] at (\segwidth,-0.25) {\textbf{$4\Delta t + \Delta$}};
\node[below] at (2*\segwidth,-0.25) {\textbf{$4\Delta t + 2 \Delta$}};
\node[below] at (3*\segwidth,-0.5) {\textbf{$4\Delta t + 3\Delta$}};
\node[below] at (4*\segwidth,-0.5) {\textbf{$4\Delta (t+1)$}};

\end{tikzpicture}
\caption{Slot $t$ of our protocol, with its four phases.}
\label{fig:pvm}
\end{figure}
\section{Analysis}
\label{sec:analysis}

Algorithm~\ref{alg:pvm-generic} works in the generalized partially synchronous sleepy model, and is in particular a $\tau$-secure ebb-and-flow protocol, \emph{if we strengthen $\tau$-compliance to require that less than $\frac{n}{3}$ validators are ever slashable for equivocation}, for reasons that will be explained shortly. For $\GST = 0$, we show in Section~\ref{sec:sync} that, if the execution is $\tau$-compliant in this stronger sense, then all the properties of $\RLMDGHOST$~\cite{rlmd} keep holding. 
In Section \ref{sec:part-sync} we show that the finalized chain $\chainfin$ is $\frac{n}{3}$-accountable, and thus always safe if $f < \frac{n}{3}$. Moreover, if $f<\frac{n}{3}$, $\chainfin$ is live after $\max\{\GST,\GAT\}$. 

Before proceeding with the analysis under synchrony and partial synchrony, we state without proof the \emph{view-merge property}, which follows from the usage of the view-merge technique, since it enables proposers to synchronize the view of honest voters with theirs. It corresponds to Lemma 2 as presented by D'Amato and Zanolini~\cite{rlmd}, with an addition regarding synchronization of the latest justified checkpoint.

\begin{lemma}
\label{thm:view-merge-property}
Suppose that $t$ is a slot with an (honest) active proposer and that network synchrony holds in rounds $[4\Delta t - \Delta, 4\Delta t + \Delta]$. Say the proposed block is $B$, and the latest justified checkpoint in the view of the proposer is $\LJ_p$. Then, at round $4\Delta t + \Delta$, all active validators broadcast a \textsc{head-vote} for the honest proposal~$B$ of slot~$t$. Moreover, $\LJ_i = \LJ_p$ for any such active validator $v_i$.
\end{lemma}

\subsection{Synchrony}
\label{sec:sync}

Throughout this part of the analysis, we assume that $\GST = 0$, and that $< \frac{n}{3}$ validators are ever slashable for equivocation, by which here we mean signing multiple \textsc{head-vote}s for a single slot, rather than violating $\mathbf{E_1}$. In other words, we are not concerned about equivocation with \textsc{ffg-vote}s, but rather with \textsc{head-vote}s, which can similarly be declared a slashable offense. Observe that, in \RLMDGHOST with fast confirmations (Appendix B~\cite{rlmd}), this assumption is strictly needed for safety (and only for clients which use fast confirmations), but for example not for reorg resilience or liveness, because fast confirmations do not affect the canonical chain. On the other hand, the protocol we present here utilizes confirmations as a prerequisite for justification, and justification does affect the canonical chain, since $\hfc$ filters out branches conflicting with the latest justified block. Therefore, we require that $< \frac{n}{3}$ validators are ever slashable for equivocation for all of the properties which we are going to prove. As already mentioned, to avoid stating it repeatedly, we further restrict $\eta$-compliant executions to those executions in which the assumption holds.

Our single slot finality protocol implemented in Algorithm~\ref{alg:pvm-generic} uses the $\hfc$ fork-choice function, dealing with checkpoints and justifications. However, one could implement it using also different fork-choice functions. In particular, we show that by substituting $\hfc$ with $\rlmdghost$ (with equal expiration period~$\eta$), \ie, if we ignore justifications and consider the \emph{vanilla} fork-choice function introduced by D'Amato and Zanolini~\cite{rlmd}, then the resulting protocol is equivalent to the \RLMDGHOST protocol with fast confirmation (Appendix B~\cite{rlmd}). This because FFG votes have no effect at all, and as such it is $\eta$-reorg-resilient, and $\eta$-dynamically-available. Moreover, the following two results about fast confirmations (Appendix B~\cite{rlmd}) also apply.

\begin{theorem}[Reorg resilience of fast confirmations]
\label{thm:fast-reorg-resilience}
Let us consider an $\eta$-compliant execution with $\GST = 0$. A block fast confirmed by an honest validator at a slot $t$ is always in the canonical chain of all active validators at rounds $\geq 4\Delta (t+1) + \Delta$.
\end{theorem}

\begin{theorem}[Liveness of fast confirmations]
\label{thm:fast-liveness}
An honest proposal $B$ from a slot $t$ in which $|H_t| \geq \frac{2}{3}n$ is fast confirmed by all active validators at round $4\Delta t + \Delta$.
\end{theorem}

We show that, under synchrony, \ie, with $\GST = 0$, these properties are preserved by our justification-respecting protocol, which uses $\hfc$ instead. To do so, we show that for every $\eta$-compliant execution, Algorithm~\ref{alg:pvm-generic} using $\FC = \rlmdghost$ and Algorithm~\ref{alg:pvm-generic} using $\FC = \hfc$ are \emph{equivalent}, i.e., the sequence of outputs of Algorithm~\ref{alg:pvm-generic} is the same regardless of which fork-choice function is used. All properties of Algorithm~\ref{alg:pvm-generic} with $\FC = \rlmdghost$ in such $\eta$-compliant executions then also apply to Algorithm~\ref{alg:pvm-generic} with $\FC = \hfc$. In particular, it is also $\eta$-reorg-resilient and $\eta$-dynamically-available, and it also satisfies reorg resilience and liveness of fast confirmations, \ie, Theorems~\ref{thm:fast-reorg-resilience} and Theorem~\ref{thm:fast-liveness} hold.

\begin{theorem}[Execution equivalence]
\label{thm:equivalence}
Let us consider an $\eta$-compliant execution with $\GST = 0$ and with Algorithm~\ref{alg:pvm-generic} using $\FC = \hfc$. Furthermore, let us consider the same execution, with the same adversarial strategy and randomness, with Algorithm~\ref{alg:pvm-generic} using $\FC = \rlmdghost$. The sequence of outputs of the two algorithms correspond exactly.
\end{theorem}

\begin{proof}
Since the only difference between the two protocols is the fork-choice function $\FC$, the sequences of outputs correspond as long as the outputs of $\hfc$ and $\rlmdghost$ obtained by active validators are always the same in the two executions. $\FC$ is used only twice in Algorithm~\ref{alg:pvm-generic}, in Line~\ref{line:alg2-fc1} for proposing, and in Lines~\ref{line:alg2-fc2}-\ref{line:alg2-fc3}, with the same input, for broadcasting a \textsc{head-vote}. We are going to prove by induction that the canonical chain of an active validator \emph{at any voting round} is the same in both executions. Since Line~\ref{line:alg2-fc2} sets $\chain_i \gets \FC(V_i, t)$, and this value is the same as in Line~\ref{line:alg2-fc3}, we only need to show that the fork-choice output in Line~\ref{line:alg2-fc1} coincides in the two executions as well. In Line~\ref{line:alg2-fc3}, an honest validator uses the fork-choice output to construct their \textsc{head-vote}s, so \textsc{head-vote}s correspond in both executions. Moreover, the view-merge property applies in both executions, so honestly proposed blocks correspond to the honest \textsc{head-vote}s from their slot. Therefore, \textsc{head-vote}s coinciding in the two executions implies that honestly proposed blocks coincide as well. Since honestly proposed blocks extend the output of the fork-choice at Line~\ref{line:alg2-fc1}, this output is then also the same in both executions, completing the proof. We now carry out the induction.

\textbf{Induction hypothesis:} At any slot $t' \leq t$ and for $r = 4\Delta t' + \Delta$, $\chain_i^r$ coincides in both executions, for any active validator $i$.

\textbf{Base case:} In rounds $[0, \Delta]$, the two executions are exactly the same, because the only justified checkpoint is \genesis, so $\hfc = \rlmdghost$. Therefore, the statement holds for $t = 0$.

\textbf{Inductive step:} Suppose now that the statement holds for~$t$, and consider round $r = 4\Delta (t+1) + \Delta$. Consider an active validator~$v_i$ with view $\V_i$ at round $r$, and latest justified block $B = LJ(\V_i)$. Let~$t'$ be minimal such that there exists a justified checkpoint~$\C = (B, t')$, \ie, slot~$t'$ is the first slot in which block~$B$ was justified. The supermajority link with target~$\C$ contains at least one FFG vote from an honest validator~$v_k$. By minimality of~$t'$, $B$ could not have been already justified in the view of~$v_k$ when broadcasting an FFG vote at slot~$t'$. Therefore, by Lines~\ref{line:set-target-checkpoint}-\ref{line:cast-ffg-vote} of Algorithm~\ref{alg:pvm-generic}, it must be the case that $B \prec \chainava_k$ at round $4\Delta t' + 2\Delta$, \ie, that it had been confirmed by~$v_k$. If it was fast confirmed at a slot $\leq t'$, then, in the execution with $\FC = \rlmdghost$, Theorem~\ref{thm:fast-reorg-resilience} implies that $B \prec \chain_j^{r'}$ for all active validators $v_j$ at any round $r' \geq 4\Delta (t'+1) + \Delta$, and so in particular that $B \prec \chain_i^r$, since $t > t'$. If instead $B \prec \chain_k^{\lceil \kappa}$ at round $4\Delta t' + 2\Delta$, \ie, $B$ is confirmed by $v_k$ due to being $\kappa$-deep in its canonical chain, then with overwhelming probability there exists a pivot slot $t'' \in [t'-\kappa, t')$ (Lemma 3~\cite{rlmd}), with proposed block~$B'$. In the execution with $\FC = \rlmdghost$, $\eta$-reorg-resilience then implies that $B' \prec \chain_j^{r'}$ for all active validators $v_j$ at any round $r' \geq 4\Delta t'' + \Delta$. In particular, $B' \prec \chain_k^{r'}$ at round $r' = 4\Delta t' + 2\Delta$, and $B' \prec \chain_i^r$. The former implies $B \prec B'$, since $B.t \leq t'-\kappa \leq B'.t$, and we then have $B \prec B' \prec \chain_i^{r}$. 

Anyway, regardless of how~$B$ has been confirmed by $v_k$, we have $B \prec \chain_i^r$. Therefore, $LJ(\V_i) = B \prec \rlmdghost(\V_i, t+1)$, which in turn implies $\rlmdghost(\V_i, t+1) = \rlmdghost \circ\,\FIL_{\text{FFG}}(\V_i, t+1) = \hfc(\V_i, t+1)$. Therefore, after~$v_i$ updates its canonical chain $\chain_i$ at round $r$ by setting $\chain_i \gets \FC(\V, t+1)$, with~$\FC$ dependent on the execution,~$\chain_i$ is the same in both executions.
\end{proof}

\subsection{Partial synchrony}
\label{sec:part-sync}

Throughout this section we assume that $f < \frac{n}{3}$. First, we prove that the finalized chain is accountably safe, exactly as done in Casper~\cite{casper}. Then, we show that honest proposals made after $\max(\GST, \GAT) + \Delta$ are justified within their proposal slot, which implies liveness of the finalized chain. 

\begin{theorem}[Accountable safety]
\label{thm:accountable-safety}
The finalized chain $\chainfin$ is accountably safe, \ie, two conflicting finalized blocks imply that at least $\frac{n}{3}$ adversarial validators can be detected to have violated either $\mathbf{E_1}$ or $\mathbf{E_2}$.
\end{theorem}

\begin{proof}
We assume throughout that there are no double justifications, \ie, there are no checkpoints $\C \neq \C'$ with $\C.t = \C'.t$, and we refer to this as the non-equivocation assumption. If that's not the case, clearly $\geq \frac{n}{3}$ validators are slashable for violating $\mathbf{E_1}$. Consider two conflicting finalized blocks $B$ and $B'$. By definition, there are also finalized checkpoints $\C$ and $\C'$ with $B = \C.B$, $B' = \C'.B$. Say $\C$ is finalized by the chain of supermajority links (\genesis, $0) \to \C_1 \dots \to \C_k = \C \to \C_{k+1}$, with $\C_{k+1}.t = \C.t + 1$, and $\C'$ by the chain (\genesis, $0) \to \C'_1 \dots \to \C'_{k'} = \C' \to \C'_{k'+1}$, with $\C'_{k'+1}.t = \C'.t + 1$.  Let $t_i = \C_i.t$, and $t'_i = \C'_i.t$. By the non-equivocation assumption, $t_k \neq t'_k$, and without loss of generality we take $t_k < t'_k$. Let $j = \min\{i\leq k'\colon t_k < t'_i\}$, so $t_k < t'_j \leq t'_{k'}$, and $t'_{j-1} \leq t_k$ by minimality of $t'_j$. By the non-equivocation assumption, $t'_j = t_{k+1}$ implies that $\C_{k+1} = \C'_j$. We then have $B = \C.B \prec \C_{k+1}.B = \C'_{j}.B \prec \C'.B = B'$, contradicting that $B$ and $B'$ are conflicting. Therefore, $t'_j > t_k + 1 = t_{k+1}$ as well. Similarly, $t'_{j-1} < t_k$. Therefore, we have $t'_{j-1} < t_k < t_{k+1} < t'_{k}$, \ie, $\C'_{j-1}.t < \C_k.t < \C_{k+1}.t < \C'_j.t$. The intersection of the two sets of voters of the supermajority links $\C_{k} \to \C_{k+1}$ and $\C'_{j-1} \to \C'_j$ contains at least $\frac{n}{3}$ validators, which are then slashable for violating $\mathbf{E_2}$. 
\end{proof}

\begin{lemma}
\label{thm:honest-slot-justifies}
\sloppy{If an honest proposer $v_p$ proposes a block $B$ at a slot $t$ after $\max(\GST,\GAT) + \Delta$}, and the latest justified checkpoint in the view of the proposer is $\LJ_p$, then the checkpoint $(B, t)$ is justified in all honest views at round $4\Delta t + 3\Delta$, by supermajority link $\LJ_p \to (B,t)$.
\end{lemma}

\begin{proof}
Since $t$ is after $\GAT + \Delta$, all $> \frac{2}{3}n$ honest validators are awake since at least round $4\Delta t - \Delta$, so at slot~$t$ they have completed the joining protocol and are active. Moreover, the view-merge property (Lemma~\ref{thm:view-merge-property}) applies to all of them. Consider now an honest validator~$v_i$. By the view-merge property, validator~$v_i$ broadcasts a \textsc{head-vote} for~$B$ at round $4\Delta t + \Delta$. Also by the view-merge property, $\LJ_i = \LJ_p$ at round $4\Delta t + \Delta$, but~$\LJ_i$ does not change until round $4\Delta t + 3\Delta$, since~$\V_i$ does not. Therefore, $\LJ_i = \LJ_p$ at round $4\Delta + 2\Delta$. By that round, all $\geq \frac{2}{3}$ honest \textsc{head-vote}s for~$B$ are received by all honest validators, including~$v_i$. Since also $B \prec \chain_i$,~$v_i$ fast confirms~$B$, and thus broadcasts an FFG vote $\LJ_i \to (B, t) = \LJ_p \to (B,t)$. All honest validators receive such votes by round $4\Delta t + 3\Delta$, and merge them into their view then. Therefore, checkpoint~$(B,t)$ is justified in all honest views at that round.
\end{proof}

\begin{theorem}[Liveness]
\label{thm:liveness}
Consider two consecutive slots $t$ and $t+1$ with honest proposers after $\max(\GST, \GAT) + 4\Delta$. The block $B$ proposed at slot $t$ is finalized at the end of slot $t+1$.
\end{theorem}

\begin{proof}
By Lemma~\ref{thm:honest-slot-justifies}, checkpoint~$(B,t)$ is justified in all honest views at round $4\Delta t + 3\Delta$. Since at the beginning of slot $t+1$ there cannot be any justified checkpoint with slot $> t$, and there cannot be any other justified checkpoint with slot $t$, $(B,t)$ is therefore the latest justified block in the view of the proposer of slot $t+1$. $B$ is then canonical in its view, and it proposes a block $B'$ which extends $B$. Again by Lemma~\ref{thm:honest-slot-justifies}, $(B', t+1)$ is justified in all honest views at round $4\Delta (t+1) + \Delta$, by the supermajority link $(B, t) \to (B',t+1)$. Therefore, $B$ is finalized in all honest views.
\end{proof}

\section{Single slot finality}
\label{sec:ssf}

The protocol implemented in Algorithm~\ref{alg:pvm-generic} is a an $\eta$-secure ebb-and-flow protocol which (at best) finalizes a block in every slot, but it does not achieve \emph{single slot finality}, \ie, it cannot finalize a proposal \emph{within its proposal slot}. At best, it lags behind by one slot, finalizing a proposal from slot $t$ at the end of slot $t+1$. A straightforward extension of our protocol which achieves single slot finality is one with $5\Delta$ rounds per slot, allowing for an additional FFG voting phase. This would be very costly in Ethereum, for two reasons. First, it would in practice significantly increase the slot time, because each voting round requires aggregating hundreds of thousands (if not millions) of BLS signatures, likely requiring a lengthier multi-step aggregation process. Moreover, it would be expensive in terms of bandwidth consumption and computation, because such votes would have to all be gossiped and verified by each validator, costly even if already aggregated. For these reasons, we describe here an alternative way to enhance to protocol for the purpose of achieving single slot finality, without suffering from the drawbacks just described. We introduce a new type of message, \emph{acknowledgment}, and a new slashing condition alongside it. Acknowledgments do not influence the protocol in any way, except in case of slashing, and are mainly intended to be consumed by external observers which want to have the earliest possible finality guarantees. Therefore, they do not need to be gossiped to and verified by all validators. They can then simply be gossiped in smaller sub-networks (similar to the \emph{attestation subnets} which Ethereum employs today), requiring limited bandwidth and verification resources. If an observer wants to have faster finality guarantees than they could have by simply following the chain or listening to the global gossip, they can opt to participate in all such sub-networks, and collect all acknowledgments. As doing so is permissionless, it can also be expected that aggregate acknowledgments, or equivalent proofs, might become available through some other channels. 

\paragraph*{Acknowledgment.}
An \emph{acknowledgment} is a tuple $[\textsc{Ack}, \C, t, v]$, where $\C$ is a checkpoint with $\C.t = t$. We also refer to this as an acknowledgment \emph{of $\C$}. A \emph{supermajority acknowledgment of $\C$} is a set of $\geq \frac{2}{3}n$ distinct acknowledgments of~$\C$. At round $4\Delta t + 3\Delta$, after merging the buffer $\mathcal{B}_i$, validator~$v_i$ broadcasts the acknowledgment $[\textsc{Ack}, \LJ_i, t, v_i]$ \emph{if $\LJ_i.t = t$}, \ie, if $\LJ_i$ has been justified in the current slot. An observer which receives a supermajority acknowledgment of a \emph{justified} checkpoint $\C$ \emph{may} consider $\C$ to be finalized.

\paragraph*{Slashing rule for finality voting.} When validator $v_i$ broadcasts an acknowledgment of~$(\C,t)$, it \emph{acknowledges} that, at the end of slot $t$, it knows about $\C$ being justified. Since the FFG voting rules prescribe that the source of an FFG vote should be the latest known justified checkpoint, subsequent FFG votes with a source whose slot is $< t$ constitute a provable violation, which is analogous to surround voting. Accordingly, we formulate a third slashing rule, which ensures that finality via a supermajority acknowledgment is accountably safe. In particular, validator~$v_i$ is slashable for an FFG vote $(\C_1, \C_2)$ and an acknowledgment $(\C, t)$, if they satisfy $\mathbf{E_3}$, \ie, $\C_1.t < \C.t < \C_2.t$. In other words, the link $\C_1 \to \C_2$ \emph{surrounds} the acknowledged checkpoint.

\begin{theorem}[Accountable safety with acknowledgments]
\label{thm:acc-saf-ack}
The finalized chain is accountably safe even when it is updated via acknowledgments as well, \ie, two conflicting finalized checkpoints imply that more than $\frac{n}{3}$ adversarial validators can be detected to have violated $\mathbf{E_1}$, $\mathbf{E_2}$, \emph{or} $\mathbf{E_3}$.
\end{theorem}

\begin{proof}
The proof largely follows that of Theorem~\ref{thm:accountable-safety}. We again consider two conflicting finalized blocks $B$ and $B'$, and corresponding finalized checkpoints $\C$ and $\C'$. Regardless of whether finalization is through a supermajority link or a supermajority acknowledgment, $\C$ and $\C'$ have to be justified, by chains of supermajority links (\genesis, $0) \to \C_1 \dots \to \C_k = \C$ and (\genesis, $0) \to \C'_1 \dots \to \C'_{k'} = \C'$. Let $t_i = \C_i.t$, and $t'_i = \C'_i.t$. By the non-equivocation assumption considered in Theorem~\ref{thm:accountable-safety}, we again have $t_k \neq t'_k$, and without loss of generality we take $t_k < t'_k$. As before, we let $j = \min\{i\leq k'\colon t_k < t'_i\}$, so $t_k < t'_j \leq t'_{k'}$, and $t'_{j-1} \leq t_k$ by minimality of $t'_j$. Moreover, also by the non-equivocation assumption, $t'_{j-1} < t_k$. If $\C$ is finalized through a supermajority link, the proof of Theorem~\ref{thm:accountable-safety} already shows that at least $\frac{n}{3}$ validators must have violated $\mathbf{E_2}$, and it is still applicable here because it does not use the last supermajority link in the chain finalizing $\C'$ (which may or may not exist here). If instead $\C$ is finalized through a supermajority acknowledgment, \ie, there are $\frac{2}{3}n$ acknowledgments of $\C$, then at least $\frac{n}{3}$ validators have violated $\mathbf{E_3}$, because $\C'_{j-1}.t < \C.t < \C'_j.t$.
\end{proof}

\begin{theorem}[Single Slot Finality]
An honest proposal from a slot $t$ after $\max(\GST, \GAT) + 4\Delta$ is finalized in round $4\Delta (t+1)$ by a supermajority acknowledgment.
\end{theorem}

\begin{proof}
Say the honestly proposed block is $B$. By Lemma~\ref{thm:honest-slot-justifies}, checkpoint $\C = (B,t)$ is justified in all honest views at round $4\Delta t + 3\Delta$. Therefore, all honest validators broadcast an acknowledgment of $\C$. Any observer which listens for acknowledgments would receive all such messages by rounds $4\Delta (t+1)$, and thus possesses a supermajority acknowledgment of $\C$. Such observer may then consider $\C$, and thus also $B$, to be finalized.
\end{proof}

\section{Conclusions}
\label{sec:conclusions}
In this work, we have made significant strides towards realizing a secure and reorg-resilient ebb-and-flow protocol that has the potential to replace Ethereum's current consensus protocol, Gasper. We have provided a comprehensive analysis and modifications to D'Amato and Zanolini's \RLMDGHOST protocol, integrating it with a partially synchronous finality gadget. In particular, our protocol introduces a novel approach for achieving single slot finality.

Another significant contribution of our work lies in the expansion of the generalized sleepy model introduced by D'Amato and Zanolini. Our generalized partially synchronous sleepy model introduces stronger constraints related to the adversary's corruption and sleepiness power and incorporates the concept of partial synchrony. This extension not only enhances the original model but also provides a generalized framework suitable for a wider array of practical scenarios.


However, despite the security guarantees of our protocol, we acknowledge that it is not (yet) practical for real-world implementation. This challenge is due to the current structure of Ethereum, which employs a large pool of validators. Requiring every validator to vote at each slot would necessitate extensive message exchanges -- a process that is far from optimal given the scale of Ethereum's network. Therefore, while our current findings represent a crucial stride towards an improved consensus protocol, they also highlight the need for additional research. Specifically, we need to focus on how we can refine the voting mechanism to better manage and aggregate the messages involved in this process.

\bibliography{references}
\bibliographystyle{plainurl}

\end{document}